\documentclass[aps,pra,reprint,amsmath,amssymb,amsfonts,twocolumn]{revtex4}   
\usepackage{graphicx,epstopdf}

\def\half{ \frac{1}{2}}

\newcommand{\be}{\begin{equation}}                                              
\newcommand{\ee}{\end{equation}}
                                                
\newcommand{\Op}[1]{\boldsymbol{\mathsf{\hat{#1}}}}
\newcommand{\vOp}[1]{\boldsymbol{\hat{\vec{\mathsf{#1}}}}}

\begin{document}                                                                

\title{Formation of deeply bound ultracold Sr$_2$ molecules 
  by photoassociation near the ${\rm ^1S+{^3P_1}}$ intercombination
  line}

\author{Wojciech Skomorowski}
\affiliation{Quantum Chemistry Laboratory, Department of Chemistry,
  University of Warsaw, Pasteura 1, 02-093 Warsaw, Poland}

\author{Robert Moszynski}
\affiliation{Quantum Chemistry Laboratory, Department of Chemistry,
  University of Warsaw, Pasteura 1, 02-093 Warsaw, Poland}

\author{Christiane P. Koch}
\email{christiane.koch@uni-kassel.de}
\affiliation{Theoretische Physik, Universit\"at Kassel,
  Heinrich-Plett-Stra{\ss}e 40, 34132 Kassel, Germany}

\begin{abstract}
We predict feasibility of the photoassociative formation of Sr$_2$
molecules in arbitrary vibrational levels of the electronic ground
state  based on state-of-the-art \textit{ab initio} calculations. 
Key is the strong spin-orbit interaction between the
c$^3\Pi_u$, A$^1\Sigma_u^+$ and  B$^1\Sigma_u^+$ states. It creates 
not only an effective dipole moment allowing free-to-bound
transitions near the ${\rm ^1S+{^3P_1}}$ intercombination line but
also facilitates 
bound-to-bound transitions via resonantly coupled excited state
%rovibrational 
levels to deeply bound %rovibrational
levels of the ground  X$^1\Sigma_g^+$ potential, with $v''$ as low as
$v''=6$.  The spin-orbit interaction is responsible for both optical
pathways. Therefore, 
those excited state levels that have the largest bound-to-bound
transition moments to deeply bound ground state levels also exhibit 
a sufficient photoassociation probability, comparable to that of the
lowest weakly bound excited state level previously observed by Zelevinsky et
al. [Phys. Rev. Lett. \textbf{96}, 203201 (2006)]. 
Our study paves
the way for an efficient photoassociative production of Sr$_2$ molecules in
ground state levels suitable for experiments testing the
electron-to-proton mass ratio.  
\end{abstract}
\maketitle

\section{Introduction}
\label{sec:intro}

The cooling and trapping of alkaline-earth metals and systems with
similar electronic structure have attracted significant attention
over the last decade.  
The interest in ultracold gases of alkaline-earth atoms
was triggered by the quest for new optical
frequency standards~\cite{LudlowSci08}. 
The extremely narrow linewidth of the intercombination  ${\rm ^1S+{^3P_1}}$  transition,
together with the magic wavelength of an optical
lattice \cite{KatoriNat05}, is at the heart of the clock proposals. 
Strontium is the atomic species of choice in many current clock
experiments~\cite{SwallowsSci11,FalkeMetro11,YamaguchiAPE11,WestergaardPRL11}.   
The narrow width of the intercombination line 
implies Doppler temperatures as low as
0.5$\,\mu$K for laser cooling~\cite{ChalonyPRL11}. It also allows for 
easy optical control of the atom-atom interactions via optical
Feshbach resonances that involve only small
losses~\cite{CiuryloPRA05,BlattPRL11}. 

The diatomic strontium molecule represents a candidate for high-precision
spectroscopy that aims at determining the  time-variation of the
electron-to-proton mass ratio~\cite{Zelevinsky:08}. The idea is to
prepare tightly confined Sr$_2$ molecules in their electronic ground
state by photoassociation in an optical lattice and carry out
high-precision Raman spectroscopy on the ground state vibrational level
spacings~\cite{Zelevinsky:08,Zelevinsky:09}. 
Photoassociation refers to the excitation of colliding atom pairs into
bound levels of an electronically excited
state~\cite{JonesRMP06}. Molecules in their electronic ground state
are obtained by spontaneous decay~\cite{FiorettiPRL98}. Whether the
excited state molecules redissociate or decay into bound ground
state levels is determined by the shape of the excited state potential
curve and possibly its coupling to other excited states. Long-range
potential wells and strong spin-orbit interaction in the excited state
of alkali dimers were found to yield significant bound-to-bound
transition matrix elements~\cite{DionPRL01}. 

To date, Sr$_2$ molecules
in their excited state have been formed by photoassociation, using
both a dipole-allowed transition~\cite{NagelPRL05,MickelsonPRL05} and
a dipole-forbidden transition near the  ${\rm ^1S+{^3P_1}}$  intercombination
line~\cite{Zelevinsky:06,Martinez08}. The formation of Sr$_2$
molecules in their electronic ground state has not yet been
demonstrated except for the very last bound level~\cite{Martinez08}. 
After photoassociation using the dipole-allowed
transition, the majority of the excited state molecules redissociates,
and only the last two bound levels of the electronic ground state can
be populated~\cite{KochPRA08b}. This is due to the long-range 
$R^{-3}$ nature of the electronically excited state (with $R$ denoting the
interatomic separation) that does not provide any mechanism
for efficient stabilization to bound ground state
levels~\cite{KochPRA08}. The situation changes for photoassociation
near the intercombination line where the excited state potential curve
in the asymptotic region behaves predominantly  as $R^{-6}$ with a small
$\delta C_3^{\textrm{res}}R^{-3}$ correction, where   $\delta C_3^{\textrm{res}}$
is proportional to $\alpha^4$ (with $\alpha$  the fine structure constant).
Large bound-to-bound transition matrix elements
with the electronic ground state that behaves asymptotically as
$R^{-6}$ are then 
expected~\cite{Zelevinsky:06}. However, quantitative estimates on
which ground state levels can be accessed were hampered to date due to
lack of reliable \textit{ab initio} information on the excited state
potential energy curves and, importantly, the spin-orbit interaction. 
The latter is crucial because it yields the effective dipole moment
that is utilized in the photoassociation transition and  also
governs possible bound-to-bound transitions following the
photoassociation. 

Here, we consider the photoassociation process of two ultracold
strontium atoms into the manifold of the coupled 
c$^3\Pi_u({\rm ^1S+{^3P}})$ + A$^1\Sigma_u^+({\rm ^1S+{^1D}})$ 
+ B$^1\Sigma_u^+({\rm ^1S+{^1P}})$
states. The excited state potential energy curves,  spin-orbit
coupling and transition dipole matrix elements are obtained by
state-of-the-art \textit{ab initio} calculations \cite{Skomorowski:12}. 
This allows us to make quantitative predictions on the photoassociation rates,
bound-to-bound transition matrix elements, and spontaneous emission
coefficients. We find that the spin-orbit interaction alters parts of
the excited state vibrational spectrum qualitatively, opening the way
for transitions into deeply bound ground state levels. This implies
that the standard picture of pure Franck-Condon type transitions near
the classical turning points in the ground and a single excited state
potential energy curve yields qualitatively wrong predictions. The
crossing between the c$^3\Pi_u({\rm ^1S+{^3P}})$ and  A$^1\Sigma_u^+({\rm ^1S+{^1D}})$
states  is  found to also significantly affect the transition
moments for the Raman spectroscopy envisioned for the test of the
electron-to-proton mass ratio. 
The paper is organized as follows: Section~\ref{sec:theory} introduces 
our model and briefly reviews the theoretical methods employed. The
numerical results are presented in
Section~\ref{sec:results}, and Section~\ref{sec:concl} concludes our paper. 

\section{Theory}
\label{sec:theory}

\begin{figure}
  \includegraphics[angle=-90,width=0.9\linewidth]{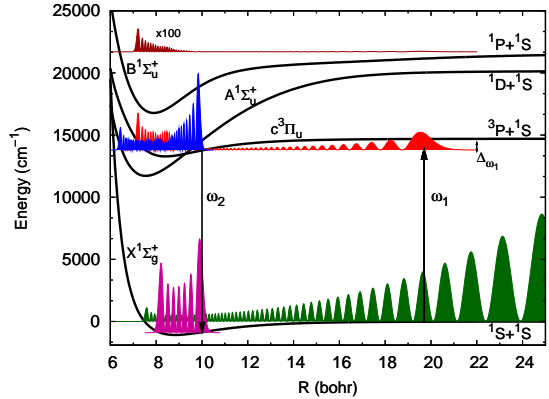}
  \caption{(Color online)
    Proposed scheme for the production of ultracold Sr$_2$
    molecules by photoassociation near the intercombination line.
    The green wavefunction represents a scattering state of two Sr
    atoms and the red, blue and brown wavefunctions the diabatic components 
    of the excited state vibrational level with binding energy
    $E_{v'=-15}=12.9\,$cm$^{-1}$.  
    Spin-orbit interaction facilitates a transition from this level to
    X$^1\Sigma_g^+$ 
    $v''=6$ (with the corresponding wavefunction depicted in purple)
    via spontaneous or stimulated emission.  
  }
  \label{fig:scheme}
\end{figure}

When a pair of colliding atoms absorbs a photon, it undergoes
a transition from the scattering continuum of the X$^1\Sigma_g^+$
ground electronic state  into  
a bound rovibrational level of an electronically excited state.
Here, we consider photoassociation using a continuous-wave
laser that is red-detuned with
respect to the $^3$P$_1$ intercombination  line of strontium.
This transition is dipole-forbidden in the nonrelativistic approximation. 
The c$^3\Pi$ state, correlating to the
asymptote of the intercombination line transition,
is, however, coupled by the spin-orbit interaction to
two singlet states, A$^1\Sigma_u^+$ and B$^1\Sigma_u^+$. 
Both singlet states are connected by a dipole-allowed transition to the
ground electronic state, $X^1\Sigma_g^+$. Thus an effective
transition matrix element is created which for moderate and large interatomic
separations is well approximated by
\begin{eqnarray}
\label{eq:dlr}
d_{{\rm SO}} &&=
\frac{\langle {\rm X}^1\Sigma_g^+|\Op{d}_{z}|{\rm B}^1\Sigma_u^+\rangle\langle {\rm B}^1\Sigma_u^+|
\Op{H}_{{\rm SO}}|{\rm c}^3\Pi_u\rangle}
{E_{{\rm c}^3\Pi_u}-E_{{\rm B}^1\Sigma_u^+}}\nonumber \\
&&+
\frac{\langle {\rm X}^1\Sigma_g^+|\Op{d}_{z}|{\rm A}^1\Sigma_u^+\rangle\langle {\rm A}^1\Sigma_u^+|\Op{H}_{{\rm SO}}|
{\rm c}^3\Pi_u\rangle}
{E_{{\rm c}^3\Pi_u}-E_{{\rm A}^1\Sigma_u^+}}\,,
\end{eqnarray}
where $\Op{H}_{{\rm SO}}$ is the spin-orbit Hamiltonian in the Breit-Pauli
approximation \cite{Bethe:57}. The long-range part of $d_{{\rm SO}}$,
dominated by the first term in the above expression, is
due to the coupling with the B$^1\Sigma_u^+$ state, ideally
suited for photoassociation. The short-range part is due to
the coupling with the A$^1\Sigma_u^+$ state, paving the way toward efficient
stabilization of the photoassociated molecules to the electronic
ground state, as we will show below. The scheme for photoassociation  
into the lowest manifold of Hund's case $(c)$ $0_u^+$ states
is depicted in Fig.~\ref{fig:scheme}.

\begin{figure}
  \includegraphics[angle=0,width=0.9\linewidth]{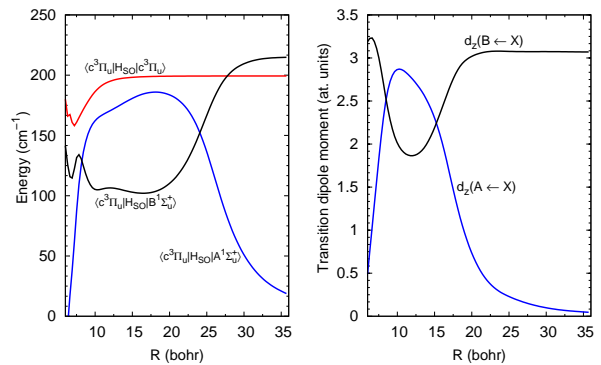}
  \caption{(Color online)
    Spin-orbit couplings (left)
    and transition dipole moments (right)
    between the relevant electronic states of the Sr$_2$ dimer
    that enter the Hamiltonian~\eqref{eq:H}.      
  }
  \label{fig:couplings}
\end{figure}
We will make use of non-adiabatic effects caused by the spin-orbit
interaction and therefore employ the diabatic (Hund's case $(a)$)
picture for our calculations. The corresponding Hamiltonian
in the rotating-wave approximation reads
\begin{widetext}
\begin{equation} \Op{{\Bbb H}}=
  \begin{pmatrix}
    \Op{H}^{{\rm X}^1\Sigma_g^+}_{\rm diag} & 0 & \half d_{z}(\rm A\leftarrow X) E_0 & 
    \half d_z(\rm B\leftarrow X) E_0  \\ 
    0 & \Op{H}^{{\rm c}^3\Pi_u^+}_{\rm diag} -A(R)-\Delta_{\omega_1} & \xi_1(R) & \xi_2(R) \\
    \half d_z(\rm A\leftarrow X) E_0 & \xi_1(R) 
    & \Op{H}^{{\rm A}^1\Sigma_u^+}_{\rm diag} -\Delta_{\omega_1}& 0 \\
    \half d_z(\rm B\leftarrow X) E_0 & \xi_2(R) & 0 
    & \Op{H}^{{\rm B}^1\Sigma_u^+}_{\rm diag} -\Delta_{\omega_1} 
  \end{pmatrix}\,,
\label{eq:H}
\end{equation}
\end{widetext}
where $d_z(n\leftarrow {\rm X})$ denotes the $z$
component of the electronic transition dipole moment from the
X electronic ground state to an electronically excited state $n$.
$R$ is the interatomic separation.
The peak amplitude and the detuning of the photoassociation
laser with respect to the intercombination line are represented by $E_0$ and
$\Delta_{\omega_1}$, respectively. 
The diagonal terms for the $(n)^{(2S+1)}|\Lambda|$ state are given by:  
\begin{widetext}
\begin{equation}
\Op{H}_{\rm diag}^{(n)^{(2S+1)}|\Lambda|}  \equiv \frac{1}{2\mu} \frac{\partial^2 } {\partial R^2} + V_{n}^{^{(2S+1)}|\Lambda|}(R)
+\frac{J(J+1)+S(S+1)-\Omega^2-\Sigma^2+L(L+1)-\Lambda^2}{2\mu
  R^2}, 
\label{eq:TT}
\end{equation}
\end{widetext}
with $\mu$ denoting the reduced mass, 
$V_{n}^{^{(2S+1)}|\Lambda|}(R)$ the radial potential energy curve, 
$J$ the rotational quantum number, and
$S$ the electronic spin quantum number. $\Lambda$, $\Sigma$, and
$\Omega$ denote the projections of the electronic orbital angular
momentum, electronic spin angular momentum, and 
the total angular momentum on the molecular axis, respectively.  
The term involving the electronic orbital quantum number $L$ in Eq.~\eqref{eq:TT}
is an approximation to the true diagonal adiabatic correction
\cite{Moszynski:06a}, with 
$L$ corresponding to the orbital quantum number in the separated-atom
limit, cf. the discussion  following  Eq. (40) in
Ref.~\cite{Moszynski:06a}. 
The spin-orbit matrix elements are defined by
\begin{eqnarray}
&&A(R)= \\
\nonumber
&& \langle {\rm c}^3\Pi_u(\Sigma=\pm1,\Lambda=\mp1)|\Op{H}_{{\rm SO}}|
{\rm c}^3\Pi_u(\Sigma=\pm1,\Lambda=\mp1)\rangle,
\end{eqnarray}
and
\begin{eqnarray}
\xi_1(R) &=&\langle {\rm c}^3\Pi_u(\Sigma=\pm1,\Lambda=\mp1)|
\Op{H}_{{\rm SO}}|{\rm A}^1\Sigma^+_u\rangle\,,\\
\xi_2(R) &=&\langle {\rm c}^3\Pi_u(\Sigma=\pm1,\Lambda=\mp1)|
\Op{H}_{{\rm SO}}|{\rm B}^1\Sigma^+_u\rangle\,.
\end{eqnarray}
The potential energy curve for the X$^1\Sigma_g^+$ ground electronic
state was taken from  Ref.~\cite{SteinPRA08}. 
All other potential energy curves, spin-orbit coupling matrix
elements (shown in the left panel of Fig.~\ref{fig:couplings}), and
electronic  transition dipole moments, $d_z(n\leftarrow {\rm X})$ 
(shown in the right panel of Fig.~\ref{fig:couplings}), 
were obtained from state-of-the-art {\em ab initio} electronic
structure calculations. The details of these calculations 
as well as their agreement with the most recent
experimental data \cite{Tiemann:11}, in particular for the crucial
A$^1\Sigma_u^+$ state,   
are reported elsewhere \cite{Skomorowski:12}.

The most promising route to form  Sr$_2$
molecules in their electronic ground state via photoassociation and
subsequent spontaneous emission is determined by 
diagonalization of the full Hamiltonian~\eqref{eq:H} and analysis of
its rovibrational structure. In order to connect our model to
experimental observables, we  calculate the photoassociation rate,
$K(\omega_1,T)$, and the branching ratios for spontaneous emission,
$P(v''\leftarrow v')$. The absorption coefficient $K(\omega_1,T)$ 
at laser frequency $\omega_1$ is given by \cite{Julienne:94,Dalgarno:71}
\begin{eqnarray}
  K(\omega_1,T)&&= \frac{2 \pi \rho^2}{\hbar Q_T}\sum_{v'J'}\sum_{J''}g_{J''}(2J''+1)
\nonumber
  \\ && \times \int_0^\infty e^{-E/k_BT}|{\cal S}_{v'J'}(E,J'',\omega_1)|^2{\rm d}E,
  \label{eq:abs1}
\end{eqnarray}
where $\rho$ denotes the gas number density, $T$ the temperature, $k_B$ the
Boltzmann constant, 
$v'$ and $J'$ the vibrational and rotational quantum numbers in the
electronically excited 
state, $J''$ the rotational quantum number of the initial scattering
state, $g_{J''}$ the spin statistical weight depending on the
nuclear spin, equal to one for $^{88}$Sr, and $Q_T=(\mu
k_BT/2\pi\hbar^2)^{3/2}$.
${\cal S}_{v'J'}(E,J'',\omega_1)$ is the ${\cal S}$-matrix element
for the transition from a continuum state with scattering energy $E$
and rotational quantum number $J''$ into the bound level $|v',J'\rangle$.
Throughout this paper, the  quantum numbers $J''$ and $v''$ 
denote the rovibrational levels of the ground electronic state,
while $J'$, $v'$ refer to the rovibrational levels
of the excited electronic state.
The square of the ${\cal S}$ matrix element in
Eq.~\eqref{eq:abs1} can be approximated by the resonant scattering
expression for an isolated resonance~\cite{Julienne:94}, 
\begin{widetext}
\begin{equation}
  |{\cal S}_{v'J'}(E,J'',\omega_1)|^2 = 
  \frac{\gamma_{v'J'}^s(E,J'')\gamma_{v'J'}^d}
  {(E-\Delta_{v'J'}(\omega_1))^2+\frac{1}{4}[\gamma_{v'}^s(E,J'')+\gamma_{v'J'}^d]^2}\,,
  \label{eq:Smat1}  
\end{equation}
\end{widetext}
where $\gamma_{v'J'}^{s}(E,J'')$ is the stimulated emission rate,
$\gamma_{v'}^{d}(E,J'')$ the rate of the spontaneous
decay, both in units of $\hbar$, $\Delta_{v'J'}(\omega_1)$ is the
detuning relative to the position of the bound rovibrational
level $ |v',J'\rangle $, i.e., $\Delta_{v'J'}=E_{v'J'}-\hbar\omega_1$, where  
$E_{v'J'}$ is the binding energy of the  level $|v',J' \rangle$. In
Eq.~\eqref{eq:Smat1}, 
we assume the decay rate due to any other undetected processes to be
negligible. 

The spontaneous emission rates
$\gamma_{v'J'}^d$  are obtained from the Einstein coefficients $A_{v'J',v''J''}$,
\begin{equation}
\gamma_{v'J'}^d=\sum_{v''J''} A_{v'J',v''J''},
\end{equation}
and related to the natural lifetimes $\tau_{v'J'}$,
$\gamma_{v'J'}^d=\hbar/\tau_{v'J'}$.
The Einstein coefficient $A_{v'J',v''J''}$  is given by 
\begin{widetext}
\begin{equation}
A_{v'J',v''J''} = \frac{4\alpha^3}{3e^4\hbar^2}
H_{J'} (E_{v'J'}-E_{v''J''})^3 
 \Big| \sum_{n'} \langle\chi^{\rm X}_{v''J''}|
d_z(n'\leftarrow {\rm X})|\chi^{n'}_{v'J'}\rangle\Big|^2\,,
\label{eq:Avv}
\end{equation}
\end{widetext}
where $H_{J'}$ is the so-called H\"onl-London factor equal to  
$(J'+1)/(2J'+1)$  for $J'=J''-1$ and $J'/(2J'+1)$ for $J'=J''+1$,
and $e$ denotes the electron charge.  
The label $n'$ represents all
considered (singlet) dissociation limits of the excited diatomic molecule,
in our case these are ${\rm ^1S\;+\;^1P}$ and  ${\rm ^1S\;+\;^1D}$. 
The non-adiabatic rovibrational wave functions
$\chi^n_{vJ}(R)=\langle R|\chi^n_{vJ}\rangle$
are obtained as the eigenfunctions of the coupled-channel Hamiltonian,
Eq.~\eqref{eq:H}, 
in the absence of the photoassociation laser field, i.e., for $E_0=0$.
In principle, in  Hund's case $(a)$,  the rovibrational wave
functions $\chi^n_{vJ}(R)$ could also be labeled,
in addition to $n$, $v$ and  $J$,
by the quantum numbers $p$, $S$, $\Sigma$, $\Lambda$ and $\Omega $, denoting the
parity, total electronic spin, its projection on the molecular axis, 
the projection of the orbital electronic angular momentum and projection 
of the total electronic angular momentum on the molecular axis
\cite{Moszynski:06a}.  
Since here we consider bosonic $^{88}$Sr atoms which 
are photoassociated to form  molecules in the rovibrational states of 
the $0_u^+$ potential, the 
parity is equal to one, and the projection of the total electronic angular 
$\Omega'$ is zero, which in turn implies $\Lambda'$=0 for singlet excited states $n'$.

At low laser intensity, $I$, the stimulated emission rate is
given by Fermi's golden rule expression:
\begin{widetext}
\begin{equation}
  \gamma_{v'J'}^s(E,J'')=4\pi^2\frac{I}{c}
    \sum_{M''=-J''}^{J''} \sum_{M'=-J'}^{J'} |
  \langle\Psi_{EJ''M''}|\vOp{d} \cdot \vec{\epsilon}|\Psi_{v'J'M'}\rangle|^2,
  \label{eq:gamma_s}  
\end{equation}
\end{widetext}
where $\vec{\epsilon}$ denotes the vector of the laser polarization,
$c$ is the speed of light,
$\Psi_{EJ''M''}$ and $\Psi_{v'J'M'}$ denote the total non-adiabatic
(electronic and rovibrational) wave functions
of the initial and final states, respectively.  $M$ is the
quantum number of the projection of the total angular momentum $J$ on
the space-fixed $Z$ axis,  
and $\vOp{d}$ denotes  the electric dipole
moment operator in the space-fixed coordinate system. 
After introducing the Born-Huang expansion of the 
non-adiabatic wave functions, Eq.~\eqref{eq:gamma_s} can
further be simplified to the following form~\cite{Moszynski:05}
\begin{eqnarray}
  \gamma_{v'J'}^s(E,J'')&&=4\pi^2\frac{I}{c}(2J'+1)H_{J'}
  \nonumber \\ && \times \Bigg|
  \sum_{n'}
  \langle\chi^{\rm X}_{EJ''}|
  d_z(n'\leftarrow {\rm X})|
  \chi^{n'}_{v'J'}\rangle\Bigg|^2,
  \label{eq:SfinBO}  
\end{eqnarray}
where $\chi^{\rm X}_{EJ''}(R)$ are energy normalized continuum
wave functions of the ground electronic state with scattering energy $E$.
Using this notation, the transition matrix elements between
coupled-channel rovibrational eigenstates become  
\begin{equation}
\langle v'',J''|d_z|v',J'\rangle \equiv \sum_{n'}
  \langle\chi^{\rm X}_{v''J''}|
  d_z(n'\leftarrow {\rm X})|
  \chi^{n'}_{v'J'}\rangle.
\end{equation}
They are almost $J-$independent as a result of the extremely small
spacings between the rotational levels of Sr$_2$. We may therefore assume
$\langle v'',J''|d_z|v',J'\rangle \approx \langle v''|d_z|v'\rangle$
(of course, the selection rule $J''=J'\pm1$ holds).

Finally, the branching ratio,
\begin{equation}
P(v''\leftarrow v'J') = \frac{\sum_{J''}A_{v'J',v''J''}}{\sum_{v''J''} A_{v'J',v''J''} }\,,
\label{eq:branch}
\end{equation}
describes the probability for the spontaneous decay from the level $|v',J'\rangle$ of the
electronically excited state to  rovibrational levels $|v'',J''=J'\pm1\rangle$ of the ground electronic 
state. Again, the branching ratio  $P(v''\leftarrow v'J')$ is nearly independent of the $J'$
quantum number.

\section{Numerical results and discussion}
\label{sec:results}

We consider $^{88}$Sr atoms trapped at a temperature
of $T\sim 2\,\mu$K, typical for the two-color mangeto-optical traps
employed for the alkaline-earth species~\cite{ZelevinskyCPC08}. 
At such a low temperature, the collisions are
purely $s$-wave, i.e., $J''=0$.
The Hamiltonian~\eqref{eq:H} is represented on a Fourier grid
with an adaptive step
size~\cite{SlavaJCP99,WillnerJCP04,ShimshonCPL06}.

\begin{figure}[tbp]
  \includegraphics[angle=-90,width=0.9\linewidth]{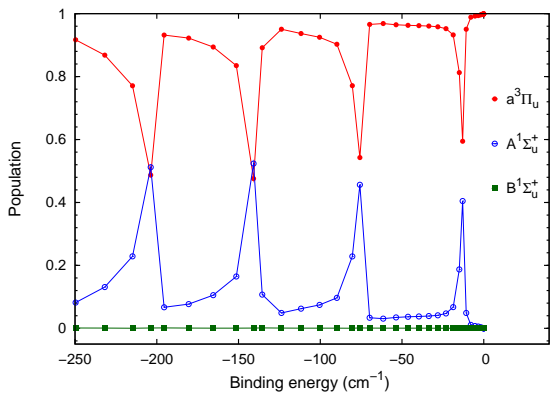}
  \caption{(Color online)
    Population of the c$^3\Pi_u$, A$^1\Sigma_u$ and B$^1\Sigma_u$
    components of the $0_u^+$ rovibrational levels for $J'=1$.
    The binding energies are taken with
    respect to the  Sr(${\rm ^3P_1}$) + Sr(${\rm ^1S}$) asymptote. 
  }
  \label{fig:pop}
\end{figure}
\begin{figure}[tbp]
\includegraphics[width=0.95\linewidth]{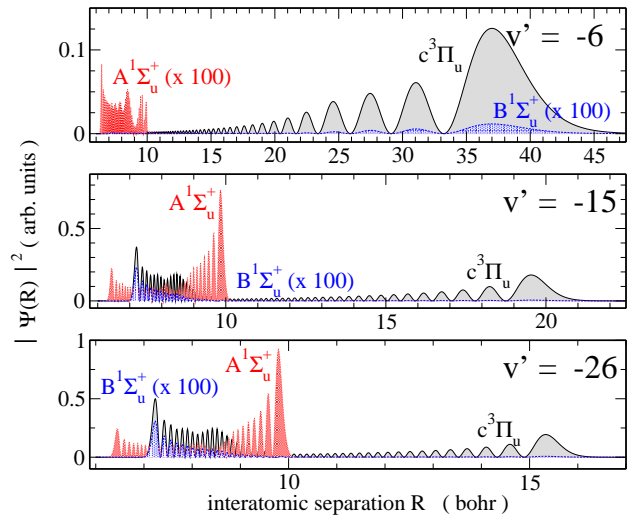}
\caption{(Color online)
  Vibrational wavefunctions of the coupled c$^3\Pi_u$, 
  A$^1\Sigma_u$ and $B^1\Sigma^+_u$ electronic states for $v'=-6$, 
  $v'=-15$  and $v'=-26$. The corresponding binding energies are 
  $E_{v'=-6}=0.27\,$cm$^{-1}=8.09\,$GHz, 
  $E_{v'=-15}=12.9\,$cm$^{-1}$, and 
  $E_{v'=-26}=75.8\,$cm$^{-1}$. Note the different scale for
  the interatomic separation. 
} 
\label{fig:vib}
\end{figure}
The photoassociation yield is determined by the ground state
scattering length and the rovibrational structure of the levels in the
excited   c$^3\Pi_u$, A$^1\Sigma_u^+$, B$^1\Sigma_u^+$ state
manifold which couple to  $0_u^+$ symmetry. 
The correct ground state scattering properties including the
scattering length are accounted for by employing the empirical X$^1\Sigma_g^+$
potential reported in Ref.~\cite{SteinPRA08}, reflecting the current
spectroscopic accuracy. The excited state rovibrational
levels are obtained from diagonalization of the Hamiltonian  \eqref{eq:H} with
$E_0=0$. Their analysis reveals a significant singlet-triplet mixing,
cf. Fig.~\ref{fig:pop} presenting the c$^3\Pi_u$,
A$^1\Sigma_u^+$ and B$^1\Sigma_u^+$ diabatic components of the coupled
wavefunctions. 
This mixing results from the crossing between the
c$^3\Pi_u$ and A$^1\Sigma_u^+$ states, which are coupled 
by spin-orbit interaction. 
On average, the rovibrational levels are predominantly of triplet
character as expected for the ${\rm ^1S\;+\;^3P}$ asymptote. 
However, a sequence of peaks indicates occurrence of
rovibrational levels with very strong singlet-triplet mixing. These
levels are particularly useful for both photoassociation
and a subsequent bound-to-bound transition. This is illustrated in 
Fig.~\ref{fig:vib} showing the vibrational wavefunctions that
correspond to the two right-most peaks in the A$^1\Sigma_u^+$ state
components of Fig.~\ref{fig:pop} (at binding 
energies of 12.9$\,$cm$^{-1}$ and 75.8$\,$cm$^{-1}$) and comparing
them to the $v'=-6$ wavefunction, the lowest level
%with $E_{v'=-6,J'=1}=0.27\,$cm$^{-1}=8.09\,$GHz, 
previously observed 
experimentally~\cite{Zelevinsky:06}. The $v'=-6$ wavefunction 
is almost purely long-range and of predominantly triplet character, with the
population of both singlet components being three orders of magnitude
smaller than the triplet one (note that the wavefunctions of both the
A$^1\Sigma_u^+$ and B$^1\Sigma_u^+$ components were scaled up by a
factor of 100 to be visible in the figure). The picture changes
completely for the levels $v'=-15$ and $v'=-26$. Since the relative 
weights of the c$^3\Pi_u$ and A$^1\Sigma_u^+$ components are almost
equal, cf. Fig.~\ref{fig:pop}, the $v'=-15$ and $v'=-26$ wavefunctions
in Fig.~\ref{fig:vib} display  A$^1\Sigma_u^+$ and c$^3\Pi_u$
components on the same scale. 
Remarkably, the triplet wavefunctions 
also show peaks at short internuclear distance.
%, close to the inner turning points of the  c$^3\Pi_u$  potential. 
This is a clear signature of resonant, non-adiabatic coupling between
vibrational levels of the spin-orbit coupled 
electronic states~\cite{AmiotPRL99,DionPRL01,GhosalKochNJP09}. It
occurs when two potential energy curves that are coupled cross and 
the energies of the two corresponding vibrational ladders 
coincide~\cite{AmiotPRL99}. Then the   
vibrational wavefunctions reflect the turning points of the two
potentials, as seen in Fig.~\ref{fig:vib}. 
Resonant coupling was shown
to lead to significantly enlarged bound-to-bound transition rates to
form deeply bound molecules in their electronic ground 
state~\cite{DionPRL01,HyewonKochPRA07,FiorettiJPhysB07}. 
According to Fig.~\ref{fig:vib}, it is the coupling between the
c$^3\Pi_u$ state and the A$^1\Sigma_u^+$ state that becomes resonant,
inducing strong mixing between these components. 
The effect of this resonant coupling will be further increased  by the
presence of the B$^1\Sigma^+_u$ state in addition to the
A$^1\Sigma_u^+$ state. The behaviour of the B$^1\Sigma^+_u$ component
strictly follows the c$^3\Pi_u$ wavefunction, but is two orders of
magnitude smaller, cf.  Fig.~\ref{fig:vib}. This is easily 
rationalized in terms of the B$^1\Sigma^+_u$ component representing
only a small admixture, due to the spin-orbit coupling $\xi_2(R)$ in the
Hamiltonian~\eqref{eq:H},  to the principal part of the $(1)0_u^+$ state
that originates  from the c$^3\Pi_u$ potential.  The magnitude of the
B$^1\Sigma^+_u$  component 
is straightforwardly estimated by treating the spin-orbit coupling as
a perturbation and calculating the  first-order correction to the
wave function, 
similarly to the expression for the transition dipole moment,
Eq.~\eqref{eq:dlr}. 

In the alkali dimers, the spin-orbit coupling mixes
in a triplet component that does not directly participate in the
optical transition between singlet
states~\cite{DionPRL01,HyewonKochPRA07,FiorettiJPhysB07}. Therefore,
the enhancement of the bound-to-bound transitions in the alkali dimers is
only due to the modification of the singlet wavefunction. Here,  
for bound-to-bound transitions to the electronic ground state, 
the effective dipole is mainly due to the coupling between the
c$^3\Pi_u$ and the A$^1\Sigma_u^+$ states, cf. Eq.~\eqref{eq:dlr}. 
Therefore, it is not only the modification of the c$^3\Pi_u$
wavefunction but also the presence of a large
A$^1\Sigma_u^+$ component that is responsible for the enhancement of
bound-to-bound transitions. Both effects together, the additional peaks
in the  c$^3\Pi_u$ wavefunction at interatomic separations
$R<10\,$bohr, and the large A$^1\Sigma_u^+$ component at these
interatomic separations lead to a significantly enhanced 
effective dipole moment according to Eq.~\eqref{eq:dlr}.
\begin{figure}[tbp]
  \includegraphics[scale=0.28,angle=-90]{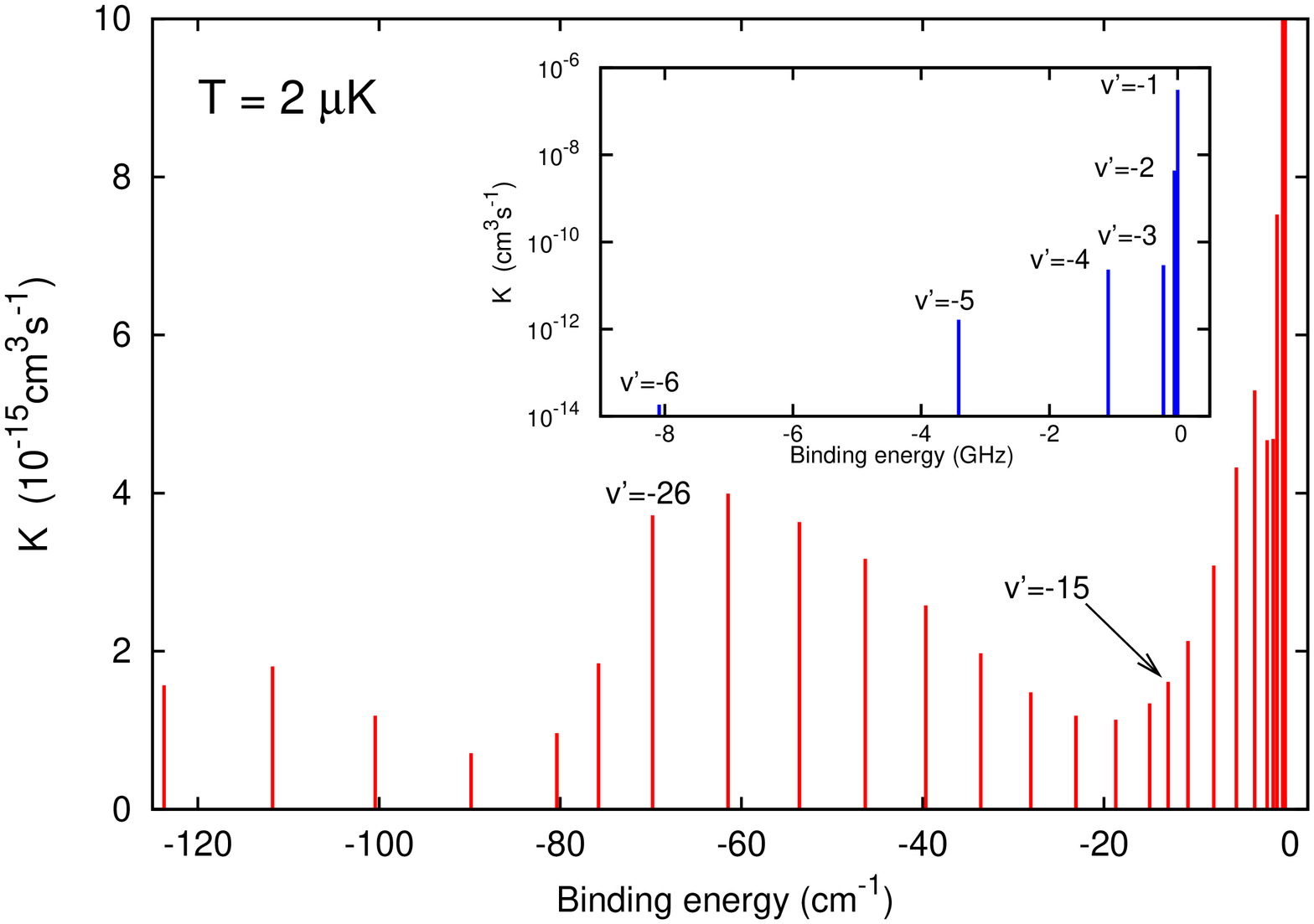}\\
  \includegraphics[scale=0.28,angle=-90]{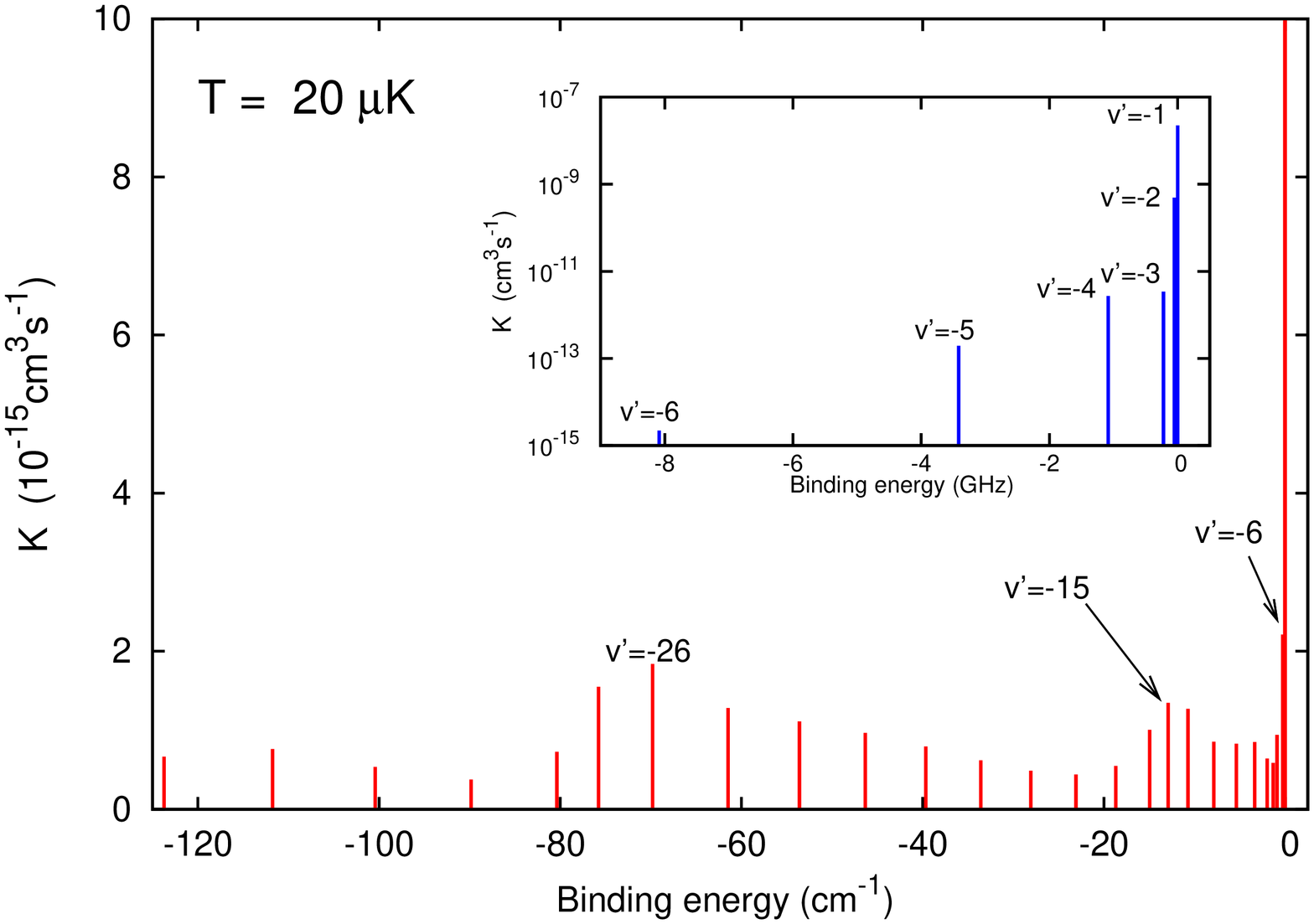}
  \caption{(Color online)
    Photoassociation into rovibrational levels of the coupled 
    c$^3\Pi_u$, A$^1\Sigma_u^+$ and B$^1\Sigma_u^+$ states below the 
    Sr(${\rm ^3P_1}$) + Sr(${\rm ^1S}$) dissociation limit 
    for a laser intensity $I=1\,$W/cm$^2$ and two temperatures,
    2$\,\mu$K (upper panel) and 20$\,\mu$K (lower panel). The
    transitions to the six least-bound levels that were reported in
    Ref.~\cite{Zelevinsky:06} are shown in a semi-logarithmic plot in
    the insert (note the different scales).
  }
  \label{fig:PAspectrum}
\end{figure}
We thus find that 
for alkaline-earth atoms near the ${\rm ^1S+{^3P_1}}$ intercombination line, the
resonant coupling enlarges the singlet admixture to a
predominantly triplet wavefunction and  enhances 
both the bound-to-bound and  the free-to-bound transition matrix elements. 
The enhancement of the bound-to-bound transitions significantly
reduces  the lifetime of the excited state bound levels. 
The lifetimes of the levels $v'=-15$ and $v'=-26$ are found to be  
30.9$\,$ns and  27.2$\,$ns, respectively, compared to 7.61 $\mu$s 
for $v'=-6$, i.e., they are decreased  by two  orders of magnitude.
This is rationalized by a larger  spontaneous emission rate resulting from 
an enhancement in the bound-to-bound transitions according to
Eq.~\eqref{eq:Avv}.

\begin{figure}[tbp]
  \includegraphics[scale=0.28,angle=-90]{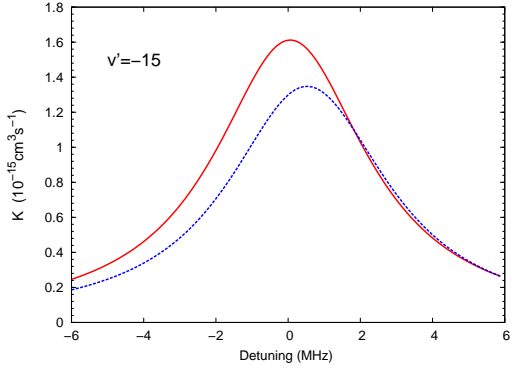}
  \caption{(Color online)
    Shape of the photoassociation line for transition into the
    rovibrational level $|v'=-15$, $J'=1\rangle$  as a function of the detuning,
    $\Delta_{v'J'}$, from  this level ($I=1$ W/cm$^2$) at $T$ = 2 $\mu$K (solid red line)
    and $T$ = 20 $\mu$K (dashed blue line).
  }
  \label{fig:PAline}
\end{figure}
The two effects, i.e., an increase in the
bound-to-bound and free-to-bound transition matrix elements,
have an opposite impact on the photoassociation probability,
with the former hindering and the latter facilitating the
photoassociation process. The photoassociation absorption coefficient,
cf. Eq.~\eqref{eq:abs1}, is shown in Fig.~\ref{fig:PAspectrum}  
for all bound levels below the ${\rm ^1S+{^3P_1}}$ dissociation limit
for two temperatures, $T=2\,\mu$K~\cite{ZelevinskyCPC08}  and 
$T=20\,\mu$K~\cite{Zelevinsky:06}. 
The absorption coefficient for the levels that were experimentally
observed~\cite{Zelevinsky:06} are shown  
in the inset of Fig.~\ref{fig:PAspectrum} using a logarithmic
scale.  At $T$ = 2 $\mu$K,
the peak rate coefficients for the strongly mixed levels
$v'=-15$ and $v'=-26$ amount to $K =1.6 \times 10^{-15}$ cm$^3$s$^{-1}$  and
$K = 1.8 \times 10^{-15}$ cm$^3$s$^{-1}$, respectively, compared to 
$K = 1.9 \times 10^{-14}$ cm$^3$s$^{-1}$ for the lowest previously
observed level, $v'=-6$, i.e., about one order of magnitude smaller. 
However, at $T$ = 20$\,\mu$K and also at higher temperatures, 
the levels with strong resonant coupling have absorption coefficients
that are very similar to that of $v'=-6$, $K = 1.3 \times
10^{-15}\,$cm$^3$s$^{-1}$ and $K=1.6 \times 10^{-15}\,$cm$^3$s$^{-1}$
for $v'=-15$ and $v'=-26$, respectively, compared to 
$K = 2.2 \times 10^{-15}$ cm$^3$s$^{-1}$ for $v'=-6$, see also bottom
panel of Fig.~\ref{fig:PAspectrum}.
The peak rate coefficients for the strongly mixed levels are less
affected by temperature broadening. This is rationalized in terms
of their large natural width, 
of the order of a few MHz. In constrast, for the level $v'=-6$  the
natural width amounts to merely 20$\,$kHz. The natural widths need to
be compared to thermal widths of 42$\,$kHz and  0.42$\,$MHz for 
$T=2\,\mu$K and $T=20\,\mu$K, respectively.  For the strongly mixed
levels, the photoassociation line shapes, shown in
Fig.~\ref{fig:PAline} for $|v'=-15$, $J'=1\rangle$,
are thus governed by
the natural width, about 4.5$\,$MHz in Fig.~\ref{fig:PAline}, 
and thermal broadening is of secondary
importance even at a temperature of $T$ = 20$\,\mu$K. 
Due to the relatively short lifetime of the level, the profile
manifests also only a very weak asymmetry~\footnote{
  Note that the detuning in Fig.~\ref{fig:PAline} is taken with respect
  to the binding energy $E_{v'=-15,J'=1}=12.9\,$cm$^{-1}$, not with
  respect to the atomic transition frequency.}.
For regular levels such as $v'=-6$ the opposite holds, i.e., the
thermal width is larger than the natural width. An increase 
in temperature from 2$\,\mu$K to  20$\,\mu$K therefore has a noticable  
effect on the photoassociation rate, cf. Ref.~\cite{CiuryloPRA04} for
a detailed analysis of the effect of thermal broadening
on the peak rate coefficients.
We conclude that photoassociation of strontium atoms into
strongly perturbed levels, albeit challenging, is within reach for 
an experimental setup such as that of Ref.~\cite{Zelevinsky:06}.

\begin{figure}[tbp]
  \includegraphics[width=0.9\linewidth]{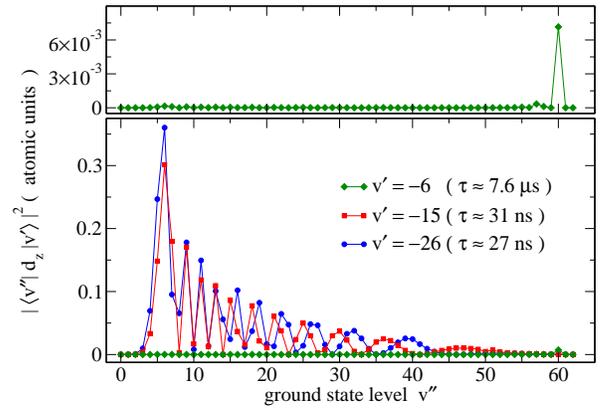}
  \caption{(Color online)
    Modulus squared of the vibrationally averaged
    bound-to-bound electric 
    transition dipole moments between excited state 
    rovibrational
    levels $v'=-6$, $v'=-15$, $v'=-26$, all with $J'=1$ (shown
    in Fig.~\ref{fig:vib}) and all vibrational levels $|v'',
    J''=0\rangle$     of the
    ground electronic state, X$^1\Sigma_g^+$. $\tau$ denotes the
    lifetime for spontaneous decay to the X$^1\Sigma_g^+$ state. 
  }
  \label{fig:FCF}
\end{figure}

\begin{figure}[tbp]
\vspace{0.3cm}
  \includegraphics[width=0.9\linewidth]{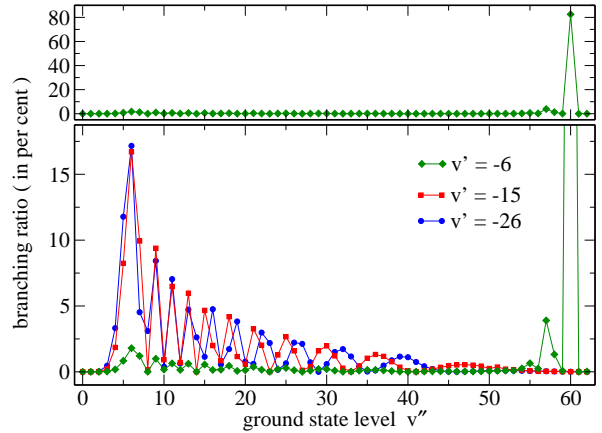}
  \caption{(Color online)
    Branching ratio for the spontaneous decay from
    levels $v'=-6$, $v'=-15$ and $v'=-26$  
    to  bound rovibrational levels of the ground electronic state.
  }
  \label{fig:branch}
\end{figure}

After observing that photoassociation into resonantly perturbed
levels such as $v'=-15$ or $v'=-26$ should be feasible
experimentally, the transition
moments from  these levels into bound levels of the electronic ground
state are examined in Fig.~\ref{fig:FCF}. Furthermore,
Fig.~\ref{fig:branch} shows the modulus squared of the
vibrationally averaged transition moments governing  the spontaneous
emission coefficients, cf. Eq.~\eqref{eq:Avv}, and the branching ratios,
cf. Eq.~\eqref{eq:branch}.
While the level $v'=-6$ decays predominantly, with a
branching ratio of more than 80\%, into $v''=-3$, a 
very weakly bound ground state level with a binding
energy of 0.17$\,$cm$^{-1}$, the strongly perturbed levels $v'=-15$ or
$v'=-26$ decay into 
a range of the ground state levels, including deeply bound ones. The
largest transition moment is observed for the ground state level
$v''=6$ with a binding 
energy of 836.4$\,$cm$^{-1}$. The corresponding branching ratios
amount to about 17\% for both  $v'=-15$ and $v'=-26$, compared to
less than 2\% for $v'=-6$. Note that the
branching ratios to $v''=6$ in Fig.~\ref{fig:branch}
are almost equal for $v'=-15$ and $v'=-26$, while the transition
moments in Fig.~\ref{fig:FCF} are not. This is due to the
dependence of the spontaneous emission coefficients on the transition
frequency in addition to the transition moment,
cf. Eq.~\eqref{eq:Avv}. Based on the favorable transition moments
between the strongly perturbed excited state levels and 
$v''=6$, stimulated emission using a nanosecond pulse could be
employed in
order to pump the excited state population selectively into the ground
state level $v''=6$. Alternatively, final state selectivity could be
achieved by photoassociation via Stimulated Raman Adiabatic Passage
(STIRAP)~\cite{MeMosheChemRev12}. It requires a sufficiently steep
trap to ensure a well-defined phase of the initial state
$|E,J''\rangle$ which is expected to be feasible in a deep optical
lattice~\cite{Tomza:11}. Due to their large transition moments for
both pump and Stokes steps, the pathways $E \to v'=-15(-26) \to
v''=6$ would be the most promising routes for STIRAP
photoassociation from an optical lattice into deeply bound levels. 

\begin{figure}[tbp]
  \includegraphics[angle=-90,width=1.05\linewidth]{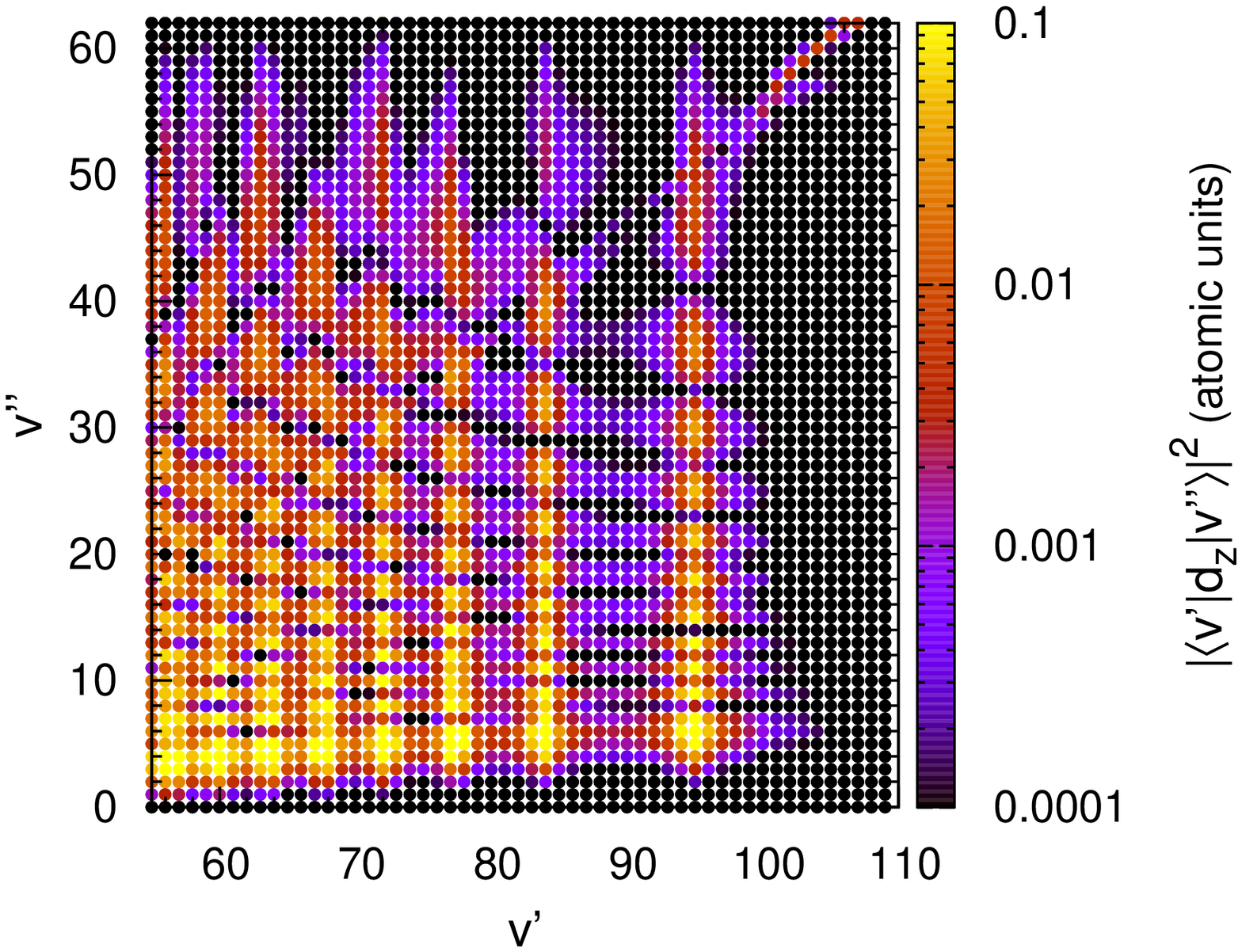}\vspace{-1cm}\\
  \includegraphics[angle=-90,width=1.05\linewidth]{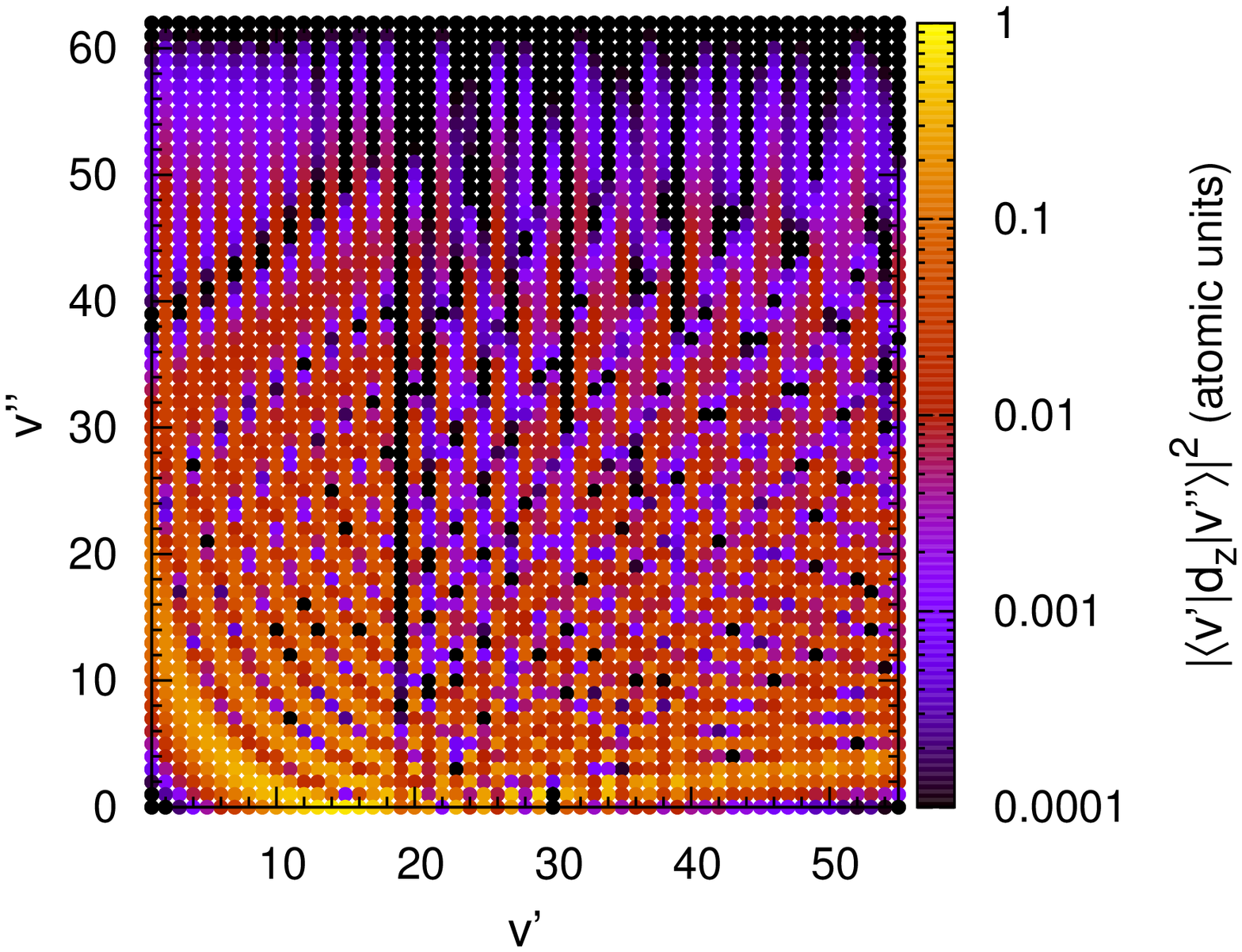}
  \caption{
    Modulus squared of the  vibrationally averaged bound-to-bound electric
    transition dipole moments between all rovibrational
    levels  $|v', J'=1\rangle$ of the $0_u^+$ potential
    and all vibrational levels $|v'', J''=0\rangle$ of the
    ground electronic state, X$^1\Sigma_g^+$
   (for other possible combinations of $J'$ and $J''$
    the pattern is almost identical). 
  }
  \label{fig:matrix}
\end{figure}
A complete overview over transitions between the rovibrational levels $v'$
of the excited $0_u^+$ states  below the ${\rm ^1S+{^3P_1}}$ asymptote
and all ground state levels $v''$ is given by
Fig.~\ref{fig:matrix}. For clarity,  
the figure has been separated into two parts, showing the highly
excited state levels $v'$ in the top  panel and 
the lower excited state levels $v'$ in the bottom panel of
Fig.~\ref{fig:matrix}. Note that we find 110 excited state
$0_u^+$ levels $v'$ below the ${\rm ^1S+{^3P_1}}$ asymptote with $J'=1$, i.e., 
$v'=-6$ corresponds to $v'=104$, $v'=-15$ to 95, 
and $v'=-26$ to 84.  Considering first levels close to 
the ${\rm ^1S+{^3P_1}}$ dissociation limit, we notice that the last
two excited state 
levels have extremely weak bound-to-bound transition moments. The next
 ten lower levels display a single peak in their transition moments,
indicating pure Franck-Condon transitions close to the outer turning
point. This is typical for weakly bound, regular levels. Transferring
the molecular population to shorter bond lengths is extremely
difficult for such levels and requires many excitation-deexcitation
cycles~\cite{MeMosheChemRev12}. 

The first strongly perturbed level,
$v'=-15$ (or $v'=95$), leads to a prominent series of peaks in the squared transition
moment matrix. Figure~\ref{fig:matrix} indicates that also the
neighbouring levels of $v'=-15$ are significantly perturbed. This
would be important for pump-dump schemes using picosecond laser
pulses~\cite{KochPRA06a,KochPRA06b}. An excited state wavepacket ideally suited
for selective population transfer into $v''=-6$ is obtained by
superimposing levels 
$v'=92,\ldots,98$. This translates into a spectral width of the
photoassociation pulse of 15$\,$cm$^{-1}$, corresponding to a
transform-limited pulse duration of 1$\,$ps. Note that a previous
study considering only the experimentally observed weakly bound levels 
concluded that short-pulse pump-dump photoassociation near the
intercombination line transition is not 
viable~\cite{KochPRA08b}. The main obstacle is the quasi-$R^{-6}$
behavior of the excited state potential that leads to a reduced
density of vibrational levels for very small photoassociation
detunings. The number of vibrational levels present is then too small
to obtain a truly non-stationary
wavepacket~\cite{KochPRA08b}. However, the picture changes completely
for more deeply bound excited state levels such as those around
$v'=95$. The spectral width of the 
pulse can easily be chosen such that several vibrational levels are
within the photoassociation window, without exciting the atomic 
intercombination line transition that  would lead to loss of
atoms~\cite{KochPRA06b}. The advantage of a time-dependent
photoassociation scheme in the presence of non-resonant coupling lies
in the dynamical interplay that arises between the interaction of the
molecule with the laser light and the spin-orbit interaction. In such
a situation,  a dynamical enhancement of the final state population
was found for strong dump pulses, indicating that 
the efficiency of population transfer is not determined
by the transition matrix elements anymore~\cite{KochPRA06b}. 

A key question is how accurate our predictions are regarding
the position of the perturbed levels such as $v'=-15$ or $v'=-26$.
There is no doubt about the presence of such levels since it results 
from the crossing between the  c$^3\Pi_u$ and  A$^1\Sigma_u^+$
potential energy curves, and this crossing was confirmed by a recent
experimental study \cite{Tiemann:11}. Our {\it ab initio} data
reported in Ref. \cite{Skomorowski:12} are able to reproduce 
the rovibrational energy levels for $J'=1$ obtained from the fit 
of the experimental data to a Dunham type expansion \cite{Tiemann:11}
to within 0.64 cm$^{-1}$. Considering all experimentally observed levels
with  $J'\le 50$, the root-mean-square deviation between
theoretically calculated levels and 
the raw experimental data is  4.5 cm$^{-1}$. 
Perhaps this value, $\pm$4.5 cm$^{-1}$, should be  considered  as 
a very conservative estimate of the error bars in the binding
energies reported in the present study. 
The main sources of error in the binding energies
are the inaccuracy of the c$^3\Pi_u$ potential and its spin-orbit correction, $A(R)$.
Scaling of the c$^3\Pi_u$ potential or the $A(R)$ coupling by  $\pm5\%$
leads to shifts  in the  binding energies by 2$\,$cm$^{-1}$ to 2.5$\,$cm$^{-1}$,  
in particular for the levels with  strong singlet-triplet mixing.
However, very good results for the A$^1\Sigma_u^+$ state
and for the atomic spin-orbit splitting of the $^3$P and $^3$D  multiplets, 
obtained in Ref. \cite{Skomorowski:12}, suggest 
the accuracy  of  the  c$^3\Pi_u$ potential and the $A(R)$ coupling
to be better than $5\%$. 
Note that scaling the other spin-orbit couplings, $\xi_1(R)$ 
and  $\xi_2(R)$, by $\pm5\%$ has a negligible effect on the position
of the bound levels. 
This confirms our assessment of  the estimated  error bars of $\pm$4.5
cm$^{-1}$ as rather conservative. While  such error bars might
appear to be relatively large  from an experimental perspective, they
are not surprising for a system with 78 electrons, strong relativistic
effects, and the  A$^1\Sigma_u^+$  potential as deep as 8433 cm$^{-1}$
that are found in the strontium dimer. 

\begin{figure}[tbp]
  \includegraphics[width=0.95\linewidth]{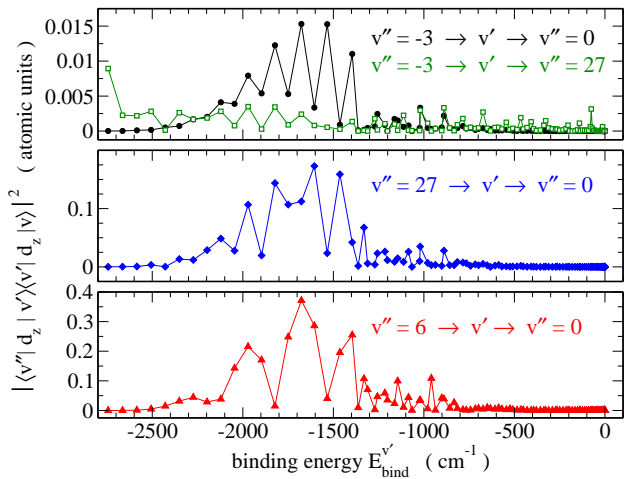}
  \caption{(Color online)
    Vibrationally averaged bound-bound Raman transition moments 
    as a function of the binding energy of the intermediate $0^+_u$
    rovibrational levels for three different pathways discussed in
    proposals for the measurement of the time variation of the
    electron-to-proton mass ratio,
    $m_e/m_p$~\cite{Zelevinsky:08,Zelevinsky:09}. 
    Note the different scales for the transition moments.
  }
  \label{fig:Raman}
\end{figure}
The Franck-Condon parabola typical for transitions between regular
vibrational levels~\cite{PilletSci08,ViteauFaraday09} is absent in 
Fig.~\ref{fig:matrix}. This reflects the strong perturbation of the 
vibrational spectrum of the excited state levels due to the spin-orbit
interaction. A reasoning on possible optical pathways solely based on
the shape of the adiabatic potentials will therefore give a wrong
picture. To emphasize this point, Fig.~\ref{fig:Raman} presents the
transition matrix elements for Raman transitions that are relevant 
in the proposal for the measurement of the time variation of the
electron-to-proton mass ratio,
$m_e/m_p$~\cite{Zelevinsky:08,Zelevinsky:09}. The idea is to transfer
molecules into the X$^1\Sigma_g^+$ ground vibrational level
starting from the weakly bound ground state level $v''=-3$
(corresponding to $v''=60$) that  
is populated by spontaneous decay from $v'=-6$, the lowest excited
state level previously observed in a photoassociation
experiment~\cite{Zelevinsky:06}.  
One could expect the efficiency of a direct
transfer $v''=-3 \to v' \to v''=0$ to be much smaller than the
efficiency of a two-Raman-step transfer employing an intermediate
state, $v''=27$. Inspection of Fig.~\ref{fig:Raman} reveals, however,
that this expectation is not confirmed. 
There exist a few excited state levels, with binding energies between
2000$\,$cm$^{-1}$ and 1400$\,$cm$^{-1}$, that have large transition
dipole moments with both $v''=-3$ (or $v''=60$) and $v''=0$, yielding a high efficiency 
for Raman transfer directly from $v''=-3$ to $v''=0$. The maximum Raman
moments are found for $v'=14$ and $v'=16$, cf. Fig.~\ref{fig:matrix}. 
These levels are almost pure singlet rovibrational states belonging to
the A$^1\Sigma_u^+$ potential, and  are only marginally  perturbed by
the spin-orbit coupling. 
For all the pathways presented in Fig.~\ref{fig:Raman} the most
favourable intermediate levels $v'$ are those which are  energetically
the highest and yet almost unperturbed, i.e., the levels located just
below the crossing between the  c$^3\Pi_u$ and  A$^1\Sigma_u^+$
potential energy curves.  This is easily rationalized in terms of 
the strong transition dipole moment,  
$d_z({\rm A \leftarrow  X})$, of these levels
and their relatively good overlap with the rovibrational levels of the
X$^1\Sigma_g^+$ potential. The decrease of the Raman transition
moments for the deeply bound levels excited state levels, with $v' \le
10$ and binding energies larger than $2000\,$cm$^{-1}$, is due to
shift of equilibrium positions of the A$^1\Sigma_u^+$ and
X$^1\Sigma_g^+$ potential wells, cf. Fig.~\ref{fig:scheme}. 

The Raman transition moments from $v''=-3 \to v'=14/16 \to v''=0$ are larger
than any of the moments for transfer from $v''=-3$ to $v''=27$. 
Of course, even higher Raman transition moments are found for optical
pathways to $v''=0$ that start in $v''=6$, cf. bottom panel of
Fig.~\ref{fig:Raman}, the level that is populated
by photoassociation into a strongly perturbed excited state level 
followed by spontaneous or stimulated emission or pump-dump
photoassociation, as explained above.  
We thus conclude that a single Raman transition after photoassociation
is sufficient to obtain molecules in the  X$^1\Sigma_g^+$ ground
vibrational level. The least intensity of the Raman lasers is required
for optical pathways starting from $v''=6$, i.e., after photoassociation
into strongly perturbed levels such as $v'=-15$ or $v'=-26$. The
pathways starting from $v''=6$
come with the additional advantage that the transition
frequencies of the Raman lasers  differ only by 
$E_{v''=0}-E_{v''=6}\approx 225\,$cm$^{-1}$ compared to
792$\,$cm$^{-1}$ for $v''=27 \to v''=0$ or 1061$\,$cm$^{-1}$ for $v''=-3
\to v''=0$.  We would like to stress here that all these conclusions 
concerning the  Raman transitions should strictly be valid 
as the intermediate $v'$ levels between 
2000$\,$cm$^{-1}$ and 1400$\,$cm$^{-1}$ are located 
in the bottom of the A$^1\Sigma_u^+$ well  where
the potential is known precisely \cite{Tiemann:11,Skomorowski:12}, 
and are  almost not perturbed by the
spin-orbit interaction. This also means that doing high-precision
Raman spectroscopy with these state should be feasible and the spectra 
will not be obscured by the spin-orbit perturbation effects.

\section{Summary and conclusions} 
\label{sec:concl}

Based on state-of-the-art \textit{ab initio} calculations, we have
calculated photoassociation rates and spontaneous emission
coefficients for the photoassociation of Sr$_2$ molecules near  the
${\rm ^1S+{^3P_1}}$ intercombination line transition. We have also
analysed bound-to-bound transition moments as well as 
Raman transition moments connecting vibrational levels in
the electronic ground state, relevant to achieve transfer into the
X$^1\Sigma_g^+$ ground vibrational level. The vibrational spectrum of
the coupled c$^3\Pi_u$, A$^1\Sigma_u^+$, B$^1\Sigma_u^+$ excited state
manifold is found to be strongly perturbed. Therefore, optical pathways cannot
be predicted based on the turning points of the adiabatic
potentials. Consequently, the theoretical analysis needs to fully account
for the spin-orbit coupling of the electronically excited states. 

For excited state binding energies of about $13\,$cm$^{-1}$ and larger,
up to $2000\,$cm$^{-1}$, strongly perturbed vibrational levels are
identified. The strong perturbations result 
from the resonant interaction of the coupled vibrational
ladders of the c$^3\Pi_u$ and A$^1\Sigma_u^+$ states. For Sr$_2$, 
these levels are found to be particularly well suited  for the 
stabilization of photoassociated molecules to the electronic ground
state, either via spontaneous or stimulated emission. 
The photoassociation rate of the strongly perturbed levels is
calculated to be comparable to that of the  lowest
level previously observed~\cite{Zelevinsky:06} 
at a temperature of $T =20\,\mu$K and about 
one order of magnitude smaller at $T =\;2$ $\mu$K. We therefore
conclude that photoassociation into strongly perturbed levels should
be feasible with the currently available experimental techniques. 

Strongly perturbed levels display 
large bound-to-bound transition
moments with deeply bound vibrational levels of the electronic ground
state. If photoassociation is followed by spontaneous emission, this
will show up as a dominant decay into X$^1\Sigma_g^+(v''=6)$, although
a large range of ground state vibrational levels will be populated as
well. State selectivity of the ground state levels can be achieved by
stimulated emission, either employing STIRAP photoassociation in a deep
optical lattice~\cite{Tomza:11} or pump-dump photoassociation with
picosecond pulses~\cite{KochPRA06a,KochPRA06b}. 

Identifying in the experiment the strongly perturbed levels of the c$^3\Pi_u$,
A$^1\Sigma_u^+$, B$^1\Sigma_u^+$ manifold that are particularly
suitable for efficient stabilization to 
deeply bound ground state levels requires a spectroscopic search since
even state-of-the-art \textit{ab inito} methods cannot predict the positions
of the rovibrational levels with precision better than a
few wavenumbers for such a heavy system like Sr$_2$. The theoretical
precision is limited here mainly by uncertainty of the c$^3\Pi_u$  
state and its relativistic correction,  and can be reduced only 
after emergence of new experimental data concerning 
the c$^3\Pi_u$, A$^1\Sigma_u^+$, B$^1\Sigma_u^+$ manifold of Sr$_2$. 

Finally, the crossing between  A$^1\Sigma_u^+$ and c$^3\Pi_u$ potentials
will be important  not only for the initial
formation of Sr$_2$ molecules but also for any subsequent Raman-type
transition proceeding via the coupled c$^3\Pi_u$, A$^1\Sigma_u^+$,
B$^1\Sigma_u^+$ manifold of states. The presence of 
unperturbed levels of the A$^1\Sigma_u^+$ state, that are
located just below  the crossing with the c$^3\Pi_u$ curve,
leads to the somewhat unexpected result that 
the weakly bound X$^1\Sigma_g^+$ vibrational levels just below the
dissociation limit show  larger Raman transition moments with the
ground vibrational level than with levels half-way down the ground state
potential well. Direct Raman transitions to the ground vibrational
level thus become possible for both weakly and strongly bound
levels. When utilizing these transitions for population transfer by 
STIRAP, deeply bound levels such as $v''=6$ come with the advantage of
a smaller frequency gap between the pump and Stokes pulse and
significantly larger transition moments translating into lower pulse
amplitudes. 

There are thus at least two  good reasons for future experiments on the
strontium dimer to employ strongly perturbed levels of the c$^3\Pi_u$,
A$^1\Sigma_u^+$, B$^1\Sigma_u^+$ manifold: efficient stabilization to 
deeply bound ground state levels and large matrix elements for
Raman transitions between ground state levels. Our calculations show
these experiments to be feasible with currently available experimental
technology. 

\section*{Acknowledgments}
We would like to thank Tanya Zelevinsky, Paul Julienne and Svetlana
Kotochigova for many useful discussions. 
This study was supported by the Polish Ministry of Science and
Higher Education through the project
N N204 215539.
Financial support from the Deutsche Forschungsgemeinschaft (Grant
No. KO 2301/2) is also
gratefully acknowledged.

%\bibliography{sr2}

\end{document}